\documentstyle[12pt]{article}
\begin{document}
\thispagestyle{empty}
\begin{center}
\LARGE \tt \bf{Non-Riemannian acoustic black hole: Hawking radiation and broken Lorentz symmetry}
\end{center}
\vspace{2.5cm}
\begin{center} {\large L.C. Garcia de Andrade\footnote{Departamento de
F\'{\i}sica Te\'{o}rica - Instituto de F\'{\i}sica - UERJ
Rua S\~{a}o Fco. Xavier 524, Rio de Janeiro, RJ
Maracan\~{a}, CEP:20550-003 , Brasil.
E-Mail.: garcia@dft.if.uerj.br}}
\end{center}
\vspace{2.0cm}
\begin{abstract}
A non-Riemannian geometrical approach to the investigation of an acoustic black hole in irrotational mean flows, based on the Lighthill vortex sound theory is given. This additional example of analog gravity based on classical fluids is used to investigate the acoustic Lorentz violation. An example is given where the contortion vector is distributed along a ring inside the fluid which can be gravitational analog of the torsion thick string spacetime defect. It is found that the linear background flow velocity approximation, acoustic Lorentz symmetry is breaking by the acoustic Cartan contortion in analogy to the spontaneous gravitational Lorentz breaking in Riemann-Cartan spacetime discovered recently by Kostelecky. Although the acoustic torsion appears in the fiducial observer acceleration it is not present in surface gravity of the acoustic black hole and we conclude that, as far as the present model is concerned, the acoustic  contortion does not contributes to Hawking radiation.
\end{abstract}
\newpage
\section{Introduction}
Recently several examples of acoustic black holes and wormholes have appeared in the literature \cite{1,2,3,4} following the earlier Hawking radiation in acoustic black hole flows discovered in 1981 by Unruh \cite{4}. Among them the comon feature of pseudo-Riemannian spacetime geometry appeared as to complete the analogy with Einstein general relativistic black holes \cite{5}. However , more recently, we discovered that the non-Riemannian geometry \cite{2} could serve as an alternative to the investigation of acoustic analog gravity as a simple way to solve some problems as the construction of a complete Riemannian geometry when one introduces the bulk vorticity in fluids as discovered by Bergliaffa et al \cite{6}. This alternative would be analogous to the Einstein-Cartan theory \cite{7} of gravity as a simple alternative of classical general relativity. In this short letter we show that is possible and necessary also to introduce Cartan torsion \cite{9} in the case of an irrotational, density inhomogeneous, mean flow which is obtained from Lighthill theory of vortex sound \cite{8}. The letter is organised as follows: In section 2 we present the non-Riemannian acoustic manifold of the irrotational inhomogeneous mean flow black hole, and also present an example of a distribution of contortion along a cylinder (not a vortex tring) inside the fluid as a gravitational analog of a spacetime defect given by a torsion string \cite{1,9}. In section 3 the generalised wave equation is solved in the eikonal approximation and the Lorentz acoustic violation is shown to possess a contribution of torsion in a special gauge in close analogy with the gravitational Lorentz violation recently discovered by Kostelecky \cite{10}. In section 4 the surface gravity is obtained and we note that the term that generates the acoustic torsion is also present in surface gravity and therefore the spectrum of Hawking radiation. Section 5 discusses the future prospects. 
\section{Non-Riemannian acoustic black holes and vortex sound}
The Lighthill theory of vortex sound has been used to investigate the complicated, and not always simple to solve analytical, problem of turbulence. Here we addopt a particular case of Lighthill vortex sound equation \cite{8}
\begin{equation}
[\frac{D}{Dt}(\frac{1}{c^{2}}\frac{D}{Dt})-\frac{1}{\rho}{\nabla}.({\rho}{\nabla})]B = \frac{1}{\rho}div({\rho}{\vec{\omega}}{\wedge}\vec{v})
\label{1}
\end{equation}
where B represents the total enthalpy \cite{10} which expression is given by
\begin{equation}
B= \int{\frac{dp}{\rho}}+\frac{1}{2}{v^{2}}
\label{2}
\end{equation}
the vortex source on the RHS of expression (\ref{1}) vanishes in irrotational regions but not everywhere ,because in this case we would have (without moving boundaries) , the enthalpy B would be constant, and there would be no sound waves in the fluid.The differential operator on the LHS of that equation represents the propagation of sound in the inhomogeneous flow. In the case of sound waves in irrotational mean flow one defines the velocity potential ${\Psi}_{0}$ where $\vec{U}={\nabla}{\Psi}_{0}$. The mean velocity of the fluid depends only on position if the fluid is bound by ,for example, the walls nozzle duct. Considering an irrotational perturbation ${\Psi}_{1}$ as
\begin{equation}
{\Psi}(\vec{x},t)= {\Psi}_{0}(\vec{x})+ {\Psi}_{1}(\vec{x},t)
\label{3}
\end{equation}
It is easy to show that this potential flow ${\Psi}$ obeys the following generalised nonlinear evolution equation
\begin{equation}
\frac{1}{c^{2}}\frac{{\partial}^{2}{\Psi}}{{\partial}t^{2}}+\frac{1}{c^{2}}\frac{D}{Dt}(\frac{1}{2}{({\nabla}{\Psi})}^{2})+\frac{1}{c^{2}}\frac{{\partial}}{{\partial}t}(\frac{1}{2}{({\nabla}{\Psi})}^{2})-{\nabla}^{2}{\Psi}=0
\label{4}
\end{equation}
where $\frac{D}{Dt}=\frac{{\partial}}{{\partial}t}+\vec{U}.{\nabla}$. According to Howe \cite{10} the linearised version of equation (\ref{4}) describes the propagation of small amplitude acoustic waves determined by ${\Psi}_{1}$. Nevertheless, when $\vec{\omega}=0$ the linearised equation for $B=-\frac{{\partial}{\Psi}_{1}}{{\partial}t}:= -\dot{\Psi}$ derived from (\ref{1}) is 
\begin{equation}
[\frac{D}{Dt}(\frac{1}{c^{2}}\frac{D}{Dt})-\frac{1}{\rho}{\nabla}.({\rho}{\nabla})]\dot{\Psi} = 0
\label{5}
\end{equation}
Since in honentropic flow the mean density and sound speed can beexpressed in terms of the variable mean velocity and the mean flow does not depend on time, we can take the perturbation potential ${\Psi}_{1}$, rather than $\dot{\Psi}$, for the acoustic variable. The lenearised equation thus becomes
\begin{equation}
[(\frac{{\partial}}{{\partial}t}+\vec{U}.{\nabla})[\frac{1}{c^{2}}(\frac{{\partial}}{{\partial}t}+\vec{U}.{\nabla})]-\frac{1}{\rho}{\nabla}.({\rho}{\nabla})]{\Psi}_{1} = 0
\label{6}
\end{equation}
where $c:=c(\vec{x})$ and ${\rho}={\rho}(\vec{x})$ are the local sound speed and density in the steady flow case. As is well known in analog gravity by comparing expression (\ref{6}) with the massless scalar field equation in Riemannian spacetime 
\begin{equation}
{\Box}{\Psi}_{1}=0
\label{7}
\end{equation}
where the ${\Box}$ is the Riemannian D'Alembertian, yields the acoustic metric  
\begin{equation}
g_{00}= \frac{\rho}{c}(U^{2}-c^{2})
\label{8}
\end{equation}
and 
\begin{equation}
g_{0i}= -\frac{\rho}{c}U_{i}
\label{9}
\end{equation}
where $(i=1,2,3)$ and the fluid is uniform in density or ${\nabla}{\rho}=0$. The other components of the acoustic metric $g_{{\mu}{\nu}}$ (where ${\mu}=1,2,3$) are $g_{ij}={\delta}_{ij}$ where ${\delta}_{ij}$ is Kr\"{o}necker delta. Collecting terms the line element becomes 
\begin{equation}
ds^{2}= \frac{\rho}{c}[(c^{2}-{\vec{U}}^{2})dt^{2}-2\vec{U}.d\vec{x}dt+dx^{2}+dy^{2}+dz^{2}]
\label{10}
\end{equation}
Note that the ergoregion $g_{00}=0$, of the acoustic black hole is determined by $c^{2}={\vec{U}}^{2}$. However, the new feature of this Letter is that 
the last term in equation $(\ref{6})$ representing the inhomogeneity in the fluid density ${\rho}$ cannot be taken into account in the acoustic Riemannian metric and therefore , as what happened in the non-Riemannian geometry of vortex flows \cite{11}, an extra non-Riemannian structure called Cartan torsion or acoustic torsion is needed. This can be better understood if one writes the massless wave equation in torsioned spacetime as
\begin{equation}
{\Box}{\Psi}_{1}= -K^{\mu}{\partial}_{\mu}{\Psi}_{1}
\label{11}
\end{equation}
where $K^{\mu}$ is the Cartan contortion vector. This equation would be the scalar massless wave equation in Riemann-Cartan spacetime. Actually comparison between the last term in equation (\ref{6}) and the RHS term in equation (\ref{11}) one obtains the expressions for the contortion components are 
\begin{equation}
\vec{K}= -\frac{{\nabla}{\rho}}{\rho}= -{\nabla}ln{\rho}
\label{12}
\end{equation}
and $K^{0}=0$ which is a comon gauge choice in torsion theories of gravity. 
Let us now consider the example of a hollow cylindrical thank of water with two heights $h_{1}=a$ and $h_{2}=b$ where $a<<b$ and consider the Heavised step function distribution in the flow density as ${\rho}={\rho}_{0}H(r-r_{0})$ where H is the step function and when $r>r_{0}$, $H=b-a$ and $H=a$ when $r<r_{0}$. Substitution of this step function density in the bounded fluid into the expression (\ref{12}) one obtains  
\begin{equation}
|\vec{K}| = \frac{{\delta}(r-r_{0})}{H(r-r_{0})}
\label{13}
\end{equation}
and at the $r=r_{0}$ infinite cylinder 
\begin{equation}
|\vec{K}| = \frac{{\delta}(r-r_{0})}{b-a}
\label{14}
\end{equation}
since the derivative of the step function is the Dirac delta distribution ${\delta}(r-r_{0})$ where $r=r_{0}$ is the radius of the cylinder inside the fluid. The acoustic contortion would then be distributed along a ring as the gravitational analog of a torsion loop spacetime defect. Note that torsion vanishes off the ring but this does not mean that there is no fluid outside the cylinder. Note that the cross section of the torsion distribution orthogonal to the z-axis represents the distribution 
\begin{equation}
|\vec{K}|{\delta}(z)= \frac{{\delta}(r-r_{0}){\delta}(z)}{H(r-r_{0})}
\label{15}
\end{equation}
which is analogous to the torsion loop distribution as a spacetime teleparallel defect \cite{9}.
\section{Acoustic Lorentz violation in non-Riemannian black holes}
In this section we shall solve the generalised wave equation $(\ref{6})$ in the case of the approximation $O({\vec{U}}^{2})=0$. This approximation is valid as long as we are outside the highly nonlinear turbulent region. Let us now consider the eikonal approximation with the ansatz
\begin{equation}
{\Psi}_{1}=a(x)e^{i({\omega}t- \vec{k}.\vec{x})}
\label{16}
\end{equation}
Expansion of expression (\ref{6}) yields
\begin{equation} 
-{\nabla}^{2}{\Psi}_{1}+\frac{1}{c^{2}}\frac{{\partial}^{2}{\Psi}_{1}}{{\partial}t^{2}}-\frac{4}{c^{3}}(\vec{U}.{\nabla}c)\frac{{\partial}{\Psi}_{1}}{{\partial}t}-({\nabla}ln{\rho}).{\nabla}{\Psi}_{1}=0
\label{17}
\end{equation}
where we drop off the term $\frac{2}{c^{3}}(\vec{U}.{\nabla}c)(\vec{U}.{\nabla}{\Psi}_{1})$ since this represents a term quadratic in the background flow velocity $\vec{U}$. In these derivations we also assume that the variation of the amplitude $a(x)$ is very slow so we do not take terms such as  ${\nabla}{a(x)}$ into consideration. It is also clear that the term ${\nabla}ln{\rho}$ would be responsible for the acoustic torsion effects in the Lorentz symmetry breaking. Substitution of expression (\ref{16}) into (\ref{17}) yields 
\begin{equation}
({\omega}^{2}-k^{2}c^{2})+\frac{4}{c}i(\vec{U}.{\nabla}c){\omega}-ic^{2}{\nabla}ln{\rho}.\vec{k}=0
\label{18}
\end{equation}
which is a second order algebraic equation in frequency ${\omega}$. Solving the equation we obtain 
\begin{equation}
{\omega}= -\frac{2}{c}i(\vec{U}.{\nabla}c)\pm\frac{1}{2}\sqrt{k^{2}+4c^{2}{\nabla}ln{\rho}-\frac{16}{c^{2}}(\vec{U}.{\nabla}c)}
\label{19}
\end{equation}
Finally we obtain the dispersion relation
\begin{equation}
{\omega}=-\frac{4}{c^{2}}(\vec{U}.{\nabla}c)[ic\pm{k}]\pm\frac{k}{2}[1+2c^{2}{\nabla}ln{\rho}.\vec{e_{k}}]
\label{20}
\end{equation}
where $\vec{e_{k}}=\frac{{\vec{k}}}{k}$ is the unit vector along the $\vec{k}$ direction. From this expression one may easily found the group velocity 
\begin{equation}
v_{g}= \frac{d{\omega}}{dk}= \mp\frac{4}{c^{2}}(\vec{U}.{\nabla}c)\pm\frac{1}{2}[1+2c^{2}{\nabla}ln{\rho}.\vec{e_{k}}]
\label{21}
\end{equation}
From expression (\ref{20}) one observes that the inhomogeneity in the phonon distribution ${\nabla}c$ yields a damping term while the acoustic torsion term contributes to the acoustic Lorentz violation. Actually this is the exact gravitational analogue of the Kostelecky spontaneous gravitational Lorentz breaking in Riemann-Cartan spacetime \cite{10}. 
\section{Hawking radiation in non-Riemannian acoustic black holes?}
To obtain the framework appropriated to derive surface gravity and Hawking radiation we write the metric (\ref{10}) for the non-Riemannian acoustic black hole in the form
\begin{equation}
ds^{2}= \frac{\rho}{c}[-c^{2}dt^{2}+ (d\vec{x}-\vec{U}dt)^{2}]
\label{22}
\end{equation} 
where the time translation Killing vector is $Q^{\mu}=(1;\vec{0})$, with norm given by 
\begin{equation}
Q^{2}= g_{{\mu}{\nu}}Q^{\mu}Q^{\nu}:= -||Q||^{2}=- \frac{\rho}{c}[c^{2}- (\vec{U})^{2}]
\label{23}
\end{equation}
By assuming that the vector $\frac{\vec{U}}{(c^{2}-{\vec{U}}^{2})}$ is integrable , a new time coordinate introduced by M. Visser \cite{3}  
\begin{equation}
d{\tau}= dt+ \frac{{\vec{U}}^{2}.d{\vec{x}}^{2}}{{c}^{2}-{\vec{U}}^{2}}
\label{24}
\end{equation}
Substitution of acoustic line element (\ref{22}) yields
\begin{equation}
ds^{2}= \frac{\rho}{c}[-(c^{2}-{\vec{U}}^{2})d{\tau}^{2}+ ({\delta}_{ij}+\frac{v_{i}v_{j}}{c^{2}-U^{2}})dx^{i}dx^{j}]
\label{25}
\end{equation} 
In this coordinate system we note that the acoustic metric is really static, where the Killing vector is hypersurface ortoghonal. With this static metric ,now one is able to obtain the surface gravity \cite{3}. Visser built this surface gravity by means of a fiducial observer (FIDOS) normalizing the time-translational Killing vector where
\begin{equation}
V_{FIDO}:= \frac{Q}{||Q||}=\frac{Q}{\sqrt{\frac{\rho}{c}[c^{2}-{\vec{U}}^{2}]}}
\label{26}
\end{equation} 
The four-acceleration is given by
\begin{equation}
A_{FIDO}:= \frac{1}{2}\frac{{\nabla}||Q||^{2}}{||Q||^{2}}
\label{27}
\end{equation} 
which can be explicitly given by \cite{3}
\begin{equation}
A_{FIDO}:= \frac{1}{2}[\frac{{\nabla}(c^{2}-{\vec{U}}^{2})}{(c^{2}-{\vec{U}}^{2})}+\frac{{\nabla}(\frac{\rho}{c})}{\frac{\rho}{c}}]
\label{28}
\end{equation} 
Due to formula (\ref{12}) one may express the fiducial acceleration in terms of the contortion vector as 
\begin{equation}
A_{FIDO}:= \frac{1}{2}[\frac{{\nabla}(c^{2}-{\vec{U}}^{2})}{(c^{2}-{\vec{U}}^{2})}-(\vec{K}+\frac{{\nabla}c}{c})]
\label{29}
\end{equation} 
However due to the conformal invariance of the surface gravity \cite{3} the density terms do not appear in its final expression 
\begin{equation}
g_{H}=\frac{1}{2}[\frac{{\partial}(c^{2}-{\vec{U}}^{2})}{{\partial}n}]
\label{30}
\end{equation} 
Thus one may conclude that the acoustic torsion is not present in the Hawking radiation as far as the present model of non-Riemannian acoustic black hole is concerned.
\section{Conclusions}
The non-Riemannian geometry of acoustic black holes were discussed in the framework of an effective Cartan contortion. An example of fluid mechanics as analog gravity model for a thick nonspinning string is presented. We note that the cross section of the spinning thick string would be the torsion loop distribution. Another example is given of the violation of acoustic torsion in analogy to the gravitational spontaneous symmetry breaking of Lorentz symmetry. These examples are built in the effective spacetime of the acoustic black hole. As shown by Visser et al \cite{11} other examples of Lorentz acoustic symmetry breaking can be provided as in the case of viscosity and in Bose-Einstein condensates. It may also possible that the acoustic torsion presents a contribution to Hawking radiation in other midels of non-Riemannian black holes as the one presenting vorticity \cite{2}. Work in this direction is now in progress. 
\section{Acknowledgments}
\paragraph{}
I am very much indebt to Dr S. Bergliaffa and Professor P.S. Letelier for enlightening discussions on the subject of this paper. Financial supports from CNPq. (Brazilian Government Agency) and Universidade do Estado do Rio de Janeiro (UERJ) are gratefully acknowledged. 

\end{document}